\documentclass[letterpaper, 10 pt, conference]{ieeeconf}  

\IEEEoverridecommandlockouts                              

\usepackage{lmodern} 
\usepackage{amsmath,amssymb,amsfonts}
\usepackage[ruled,vlined]{algorithm2e}
\usepackage{graphicx}
\usepackage{textcomp}
\usepackage{xcolor}
\usepackage{booktabs} 
\usepackage{listings}
\usepackage{subcaption}
\usepackage{multirow}
\usepackage{balance}
\usepackage[super]{nth}
\usepackage[switch, mathlines]{lineno}
\usepackage{float}
\usepackage{url}
\usepackage[multiple]{footmisc}
\usepackage[export]{adjustbox}
\usepackage{xspace}
\usepackage{tcolorbox}
\usepackage{array}
\usepackage{gensymb}
\usepackage{tabularx}
\usepackage{longtable}
\usepackage{supertabular,booktabs}
\usepackage{hhline}
\usepackage{colortbl}
\usepackage{multirow}
\usepackage[short]{optidef}

\usepackage{acronym}
\newcommand{\nox}{$\mathrm{NO_{x}}$~}

\newcommand{\norm}[1]{\left\lVert#1\right\rVert}
\newcommand{\trans}{^\top}
\newcommand{\yimep}[1][]{y_{\text{IMEP}#1}}
\newcommand{\ynox}[1][]{y_{\text{NO}_x#1}}
\newcommand{\yca}[1][]{y_{\text{CA50}#1}}
\newcommand{\ymprr}[1][]{y_{\text{MPRR}#1}}
\newcommand{\ufuel}[1][]{u_{\text{DOI,fuel}#1}}
\newcommand{\uwater}[1][]{u_{\text{DOI,water}#1}}
\newcommand{\unvo}[1][]{u_{\text{NVO}#1}}

\newcommand{\rimep}[1][]{r_{\text{IMEP}#1}}

\usepackage{tikz}	
\usepackage{pgfplots}                              
\usepackage{pgfgantt}
\usepackage{pdflscape}
\usepackage{standalone}
\usetikzlibrary{arrows,shapes,positioning,shadows,trees}
\usetikzlibrary{plotmarks}
\usetikzlibrary{calc,shapes.multipart,shapes.arrows}
\usepackage{qtree}
\pgfplotsset{compat=newest}
\pgfplotsset{plot coordinates/math parser=false}

\tikzset{
	papDecision/.style = {
		diamond,
		draw, 
		text width = 18 mm, 
		align = center, 
		text badly centered,
		inner sep = 1 pt,
		font=\footnotesize,
		minimum width = 25mm,
		minimum height = 7mm,
	},
	papStart/.style = {
		rectangle,
		draw, 
		align = center, 
		text width = 3.3cm, 
		text badly centered,
		inner sep = 4 pt,
		rounded corners=10pt,
		font=\footnotesize,
		minimum width = 30mm,
		minimum height = 7mm,
	},
	papEnd/.style = {
		rectangle,
		draw, 
		align = center, 
		text width = 3.3cm, 
		text badly centered,
		inner sep = 4 pt,
		rounded corners=10pt,
		font=\footnotesize,
		minimum width = 30mm,
		minimum height = 7mm,
	},
	papData/.style = {
		trapezium,
		draw, 
		align = center, 
		text width = 20 mm, 
		text badly centered,
		inner sep = 4 pt,
		trapezium left angle=70,
		trapezium right angle=110,
		font=\footnotesize,
		minimum width = 30mm,
		minimum height = 7mm,
	},
	papPredProc/.style = {
		draw,
		rectangle split,
		rectangle split horizontal,
		rectangle split parts = 3,
		rectangle split empty part width=-8pt,
		align = center, 
		text badly centered,
		font=\footnotesize,
		minimum width = 30mm,
		minimum height = 7mm,
	},
	papProcess/.style = {
		rectangle,
		draw,
		align = center, 
		text width = 3.3cm, 
		text badly centered,
		font=\footnotesize,
		minimum width = 30mm,
		minimum height = 7mm,
	},
	papLine/.style = {
		draw,
		-stealth,
		font=\footnotesize,
	},
}

\newlength\figH
\newlength\figW

\usepackage[backend=bibtex,sorting=none,style=numeric-comp, citestyle=numeric-comp,doi=false,url=false]{biblatex}
\bibliography{BIB_dgordon.bib} 

\AtEveryBibitem{\clearfield{issn}}
\AtEveryCitekey{\clearfield{issn}}

\title{\LARGE \bf
Introducing a Deep Neural Network-based Model Predictive Control Framework for Rapid Controller Implementation}
\author{David C. Gordon $^{1}$, Alexander Winkler $^{1}$, Julian Bedei $^{1}$, \\Patrick Schaber $^{1}$, Jakob Andert $^{1}$ and Charles R. Koch $^{2}$
\thanks{$^{1}$Teaching and Research Area Mechatronics in Mobile Propulsion, RWTH Aachen University, Forckenbeckstrasse 4, 52074 Aachen, Germany. andert@mmp.rwth-aachen.de}%
\thanks{$^{2}$University of Alberta, 116 St and 85 Ave, Edmonton, AB T6G 2R3, Canada. dgordon@ualberta.ca}%
}

\begin{document}

\maketitle
\thispagestyle{empty}
\pagestyle{empty}

\begin{abstract}
Model Predictive Control (MPC) provides an optimal control solution based on a cost function while allowing for the implementation of process constraints. As a model-based optimal control technique, the performance of MPC strongly depends on the model used where a trade-off between model computation time and prediction performance exists. One solution is the integration of MPC with a machine learning (ML) based process model which are quick to evaluate online. This work presents the experimental implementation of a deep neural network (DNN) based nonlinear MPC for Homogeneous Charge Compression Ignition (HCCI) combustion control. The DNN model consists of a Long Short-Term Memory (LSTM) network surrounded by fully connected layers which was trained using experimental engine data and showed acceptable prediction performance with under 5\% error for all outputs. Using this model, the MPC is designed to track the Indicated Mean Effective Pressure (IMEP) and combustion phasing trajectories, while minimizing several parameters. Using the \texttt{acados} software package to enable the real-time implementation of the MPC on an ARM Cortex A72, the optimization calculations are completed within 1.4 ms. The external A72 processor is integrated with the prototyping engine controller using a UDP connection allowing for rapid experimental deployment of the NMPC. The IMEP trajectory following of the developed controller was excellent, with a root-mean-square error of 0.133 bar, in addition to observing process constraints. 


\end{abstract}

\section{Introduction}

Model-based optimal control techniques leverage the significant advances in system modeling and the computational performance increases seen over the last two decades~\cite{Rawlings2017}. A wide range of model-based control methods have been investigated including: linear quadratic regulator~\cite{lopez2011lqr}, sliding mode controller~\cite{norouzi2020adaptive}, adaptive control~\cite{lavretsky2013robust}, and Model Predictive Control~(MPC)~\cite{ICEMPC8}. Of~these model-based control strategies MPC is the most widely used in a range of applications~\cite{lee2011model}. Which takes advantage of the ability of MPC to provide an optimal control solution while allowing for the implementation of constraints on system states and controller outputs~\cite{AllgNMPCandMHE1999}. For automotive applications, MPC sounds like an ideal solution as internal combustion engines (ICEs) have nonlinear process dynamics and have many constraints that must be obeyed, however, the high computational demand of the online MPC has meant that it has only recently been applied to vehicles~\cite{liao2020model,norouzi2023integrating}. One such solution is the integration of black-box machine learning (ML) process models which are quick to evaluate during MPC optimization.

One specific application of interest for the application of MPC is the combustion control of Homogeneous Charge Compression Ignition~(HCCI) which has shown promise in reducing engine-out emissions (99\% reduction in \nox vs modern gasoline combustion~\cite{Breitbach.2013}) and increasing efficiency (up to 30\% compared to current gasoline engines~\cite{AdcockSAE2017}). The challenge is that HCCI is susceptible to large cyclic variations and correspondingly poor combustion stability as the combustion process lacks a direct mechanism to control combustion timing~\cite{fathi2017}. Specifically, HCCI operation suffers from significant cycle-to-cycle coupling, resulting from high exhaust gas re-circulation, leading to bounded operation due to misfire at low loads and high-pressure rise rates and peak pressure at high loads which complicates modeling and control implementation. 

Several ML techniques have been widely used for addressing engine performance, emission modeling and control~\cite{aliramezani2022modeling, Yu2018,bedei2023dynamic}. As HCCI combustion experiences significant cyclic coupling a Recurrent Neural Network (RNN) is of interest for process modeling due to the inclusion of backward connections to handle sequential inputs~\cite{Geron2019}. However, the contribution of earlier time steps become increasingly small (the ``vanishing gradient'') and thus RNN cannot accurately capture long-term dependencies in the process. Therefore, memory cells can be introduced of which the Long Short-Term Memory (LSTM) cell is the most common~\cite{Hochreiter.1997}. Each LSTM cell has two recurrent loops, one for long-term information and one for short-term information. However, traditionally the integration of an LSTM model into a nonlinear Model Predictive Controller (NMPC) has been focused on slow-response applications such as temperature set-point planning for buildings~\cite{Jeon2021}. Recent progress in the experimental application of LSTM-NMPC to high-speed systems including diesel combustion control has shown great potential of the strategy~\cite{GordonEnergies2022}. Unlike the existing literature, this work demonstrates the implementation of a real-time controller on an external ARM Cortex~A72 processor using the \texttt{acados} optimal control framework to allow for real-time execution of the NMPC for control of HCCI combustion~\cite{Verschueren2019}. The controller will be designed to track engine load and combustion timing trajectories while minimizing fuel and water consumption, pressure rise rates, and \nox emissions.

\section{Experimental Setup}
 A single cylinder research engine~(SCRE) outfitted with a fully variable electromechanical valve train~(EMVT) is used. The EMVT system allows for engine operation with a variety of valve strategies, however, in this work only negative valve overlap~(NVO) will be used to provide the required thermal energy for HCCI combustion. Fuel, conventional European RON~95 gasoline with 10~\% ethanol, and distilled water is directly injected into the combustion chamber. Full engine details can be found in~\cite{Gordon_2023}. The SCRE is controlled using a dSPACE MicroAutoBox II (MABX) rapid control prototyping (RCP) ECU containing an Xilinx Kintex-7 FPGA which is used to calculate combustion metrics in real-time~\cite{Gordon2018}.

\section{Deep Neural Network-based Engine Model}
\begin{figure*}
	\centering
	\includegraphics[trim = 0 25 0 0, clip, width = 1\textwidth]{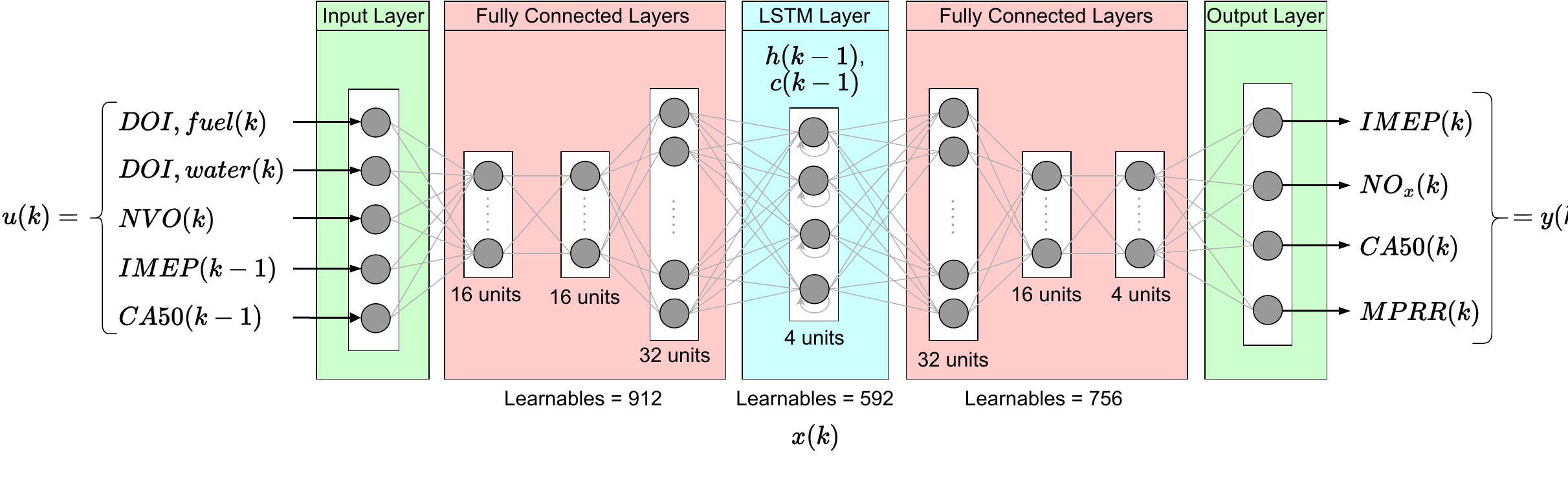}
	\caption[Structure of proposed deep neural network model for engine performance and emission modeling.]{Structure of proposed deep neural network model for engine performance and emission modeling. LSTM: Long short-term memory, DOI: duration of injection, IMEP: indicated mean effective pressure, MPRR: maximum pressure rise rate, CA50: combustion phasing for 50\% heat release
		\label{fig:lstmmodelstructure}}
\end{figure*}

To model the HCCI engine performance and emissions, a deep neural network~(DNN) with seven hidden layers (six Fully Connected (FC) layers and one LSTM layer) was developed as shown in Fig~\ref{fig:lstmmodelstructure}.

To train this DNN, which has 2260 learnable parameters, a data set of 65,000 consecutive cycles was collected from the SCRE. During the engine operation, the process inputs of duration of injection (DOI) of fuel, DOI of water, and negative valve overlap (NVO) duration were varied in both amplitude and frequency. Then, using a similar training process to previous work, the combustion outputs of engine load represented by indicated mean effective pressure (IMEP), combustion phasing angle (CA50), maximum pressure rise rate (MPRR) and engine out nitrogen oxide emissions \nox were predicted~\cite{GordonEnergies2022}. The overall accuracy of this model for each output is summarized in Table~\ref{tab:lstmtrainingvalidation} which lists the mean squared error in percent for training and validation data. MPRR was the most difficult parameter to predict, with a 4.7\% error on the validation dataset, while other outputs were predicted with less than 4\% error. 

\begin{table}[htb]
	\centering
	\caption{RMSE and normalized RMSE of DNN model vs. experimental data} \label{tab:lstmtrainingvalidation}
	\begin{tabular}{lccc}
		\hline
		& \textbf{Unit} & \textbf{Training}&\textbf{Validation}\\
		\hline
		\multirow{2}{*}{$\yimep$} & [bar] & 0.074 & 0.077 \\
		& [\%] & 2.7 & 2.8 \\ 
		\hline
		\multirow{2}{*}{$\ynox$} & [ppm] & 18 & 16 \\
		& [\%] & 4.2 & 3.8 \\ 
		\hline
		\multirow{2}{*}{$\ymprr$} & [bar/CAD] & 1.6 & 1.7 \\
		& [\%] & 4.4 & 4.7 \\ 
		\hline
		\multirow{2}{*}{$\yca$} & [CAD] & 1.5 & 1.6 \\
		& [\%] & 3.4 & 3.6 \\ 
		\hline
	\end{tabular}
\end{table}
For implementation into an MPC, this model is formulated using a nonlinear state-space representation to allow for integration into \texttt{acados} as described in~\cite{Norouzi2022}. The nonlinear state-space model is given by
\begin{subequations}
	\begin{align}
		x(k+1) &= f\left(x(k), u(k)\right),
		\\
		\begin{split}
			{y(k)} &= f_\text{FC,out}\left(f\left(x(k), u(k)\right)\right),
			\\
			&= g\left(x(k), u(k)\right),
		\end{split}
	\end{align}
\end{subequations}
with $f$ combining the input and LSTM layers, $f_{FC,out}$ representing the output layers. Here $x(k)$ are the internal model states (corresponding to the LSTM: four cell states, $c(k)$, and four hidden states, $h(k)$), ${y(k)}$ the model outputs, and $u(k)$ the model inputs. These are: 
\label{eq:ss_final}
\begin{align}
		x(k) &= \begin{bmatrix} c(k-1) \\ h(k-1) \end{bmatrix} \in\mathbb{R}^8, \quad 
		y(k) = \begin{bmatrix} \yimep(k)  \\ \yca(k) \\ \ynox(k) \\ \ymprr(k) \end{bmatrix} \in\mathbb{R}^4,
		\nonumber \\
		u(k) &= \begin{bmatrix} \yimep(k-1) \\ \yca(k-1) \\ \ufuel(k) \\ \uwater(k) \\ \unvo(k) \end{bmatrix}\in\mathbb{R}^5.
\end{align}


To reduce the oscillation of the outputs, the gradient of manipulated variables is added as new inputs~\cite{Diehl2009}. This ensures that the positive definite weighting matrix forces the change to zero thus allowing the controller to achieve the desired output setpoints:
\begin{subequations}
\label{eq:ss_final}
    \begin{align}
    \underbrace{\begin{bmatrix} x(k+1) \\ u(k) \end{bmatrix}}_{\tilde{x}(k+1)} &= \underbrace{\begin{bmatrix} f(x(k), u(k-1) + \Delta u(k)) \\ u(k-1) + \Delta u(k) \end{bmatrix}}_{\tilde{f}(\tilde{x}(k), \Delta{u}(k))},
    \\
    \underbrace{\begin{bmatrix} y(k) \\ u(k-1) \end{bmatrix}}_{\tilde{y}(k)} &= \underbrace{\begin{bmatrix} g(x(k)) \\ u(k-1) \end{bmatrix}}_{\tilde{g}(\tilde{x}(k))}.
    \end{align}
\end{subequations}
This formulation allows for both the absolute inputs and their gradient to be penalized within the cost function.

Thus, the discrete Optimal Control Problem (OCP) is defined as follows
\begin{mini}
	{\substack{\Delta{u}_0, \dots, \Delta{u}_{N-1} \\ \tilde{x}_0, \dots, \tilde{x}_N \\ \tilde{y}_0, \dots, \tilde{y}_N}}%
	{\sum_{i=0}^{N} \norm{r_i-\tilde{y}_i}^2_Q + \norm{\Delta{u}_j}^2_R}
	{}{}
	\addConstraint{\tilde{x}_0}{=\begin{bmatrix}x(k),~u(k-1)\end{bmatrix}\trans}
	\addConstraint{\tilde{x}_{i+1}}{= \tilde{f}(\tilde{x}_i, \Delta{u}_i)}{\quad\forall i\in\mathbb{H}\setminus N}
	\addConstraint{\tilde{y}_i}{= \tilde{g}(\tilde{x}_i, \Delta{u}_i)}{\quad\forall i\in\mathbb{H}}
	\addConstraint{u_\text{min}}{\le F_u \cdot \tilde{u}_k \le u_\text{max}}{\quad\forall i\in\mathbb{H}}
	\addConstraint{y_\text{min}}{\le F_y \cdot \tilde{y}_k \le y_\text{max}}{\quad\forall i\in\mathbb{H}}
	\label{eq:ocp}
\end{mini}
where $\mathbb{H}=\left\{0, 1, \dots, N\right\}$.
The reference $\tilde{r}_i$ and the weighting matrix $Q$ are selected such that deviations from the requested IMEP and CA50 are penalized while minimizing \nox~emissions, duration of injected fuel $\text{DOI,fuel}(k)$ and water $\text{DOI,water}(k)$ and change in control input $\Delta{u}$. Therefore, the specific cost function $J$ is specified as
\begin{align}
	\label{eqn:cost_nmpc}
	J &= \sum_{i=0}^{N} \underbrace{\norm{\rimep[,i] - \yimep[,i]}^2_{q_\text{IMEP}}  }_{\text{Reference Tracking}} \nonumber \\
     &+ 
    \underbrace{\norm{r_{\text{CA50,i}} - \yca[,i]}^2_{q_\text{CA50}} }_{\text{Reference Tracking}} \nonumber \\	
     &+    \underbrace{\norm{\ufuel[,i]}^2_{r_\text{DOI,fuel}} +\norm{\uwater[,i]}^2_{r_\text{DOI,water}}}_{\text{Fuel / Water consumption reduction}} \nonumber \\
     &+      \underbrace{\norm{\ynox[,i]}^2_{q_{\text{NO}_x}}}_{\text{Emission Reduction}}  +    \underbrace{\norm{\Delta{u}_i}^2_R}_{\text{Oscillation Reduction}}	 
 \end{align}
One significant advantage of using MPC for combustion control is the ability to impose constraints on inputs and outputs to ensure safe engine operation. $F_u$ and $F_y$ in Eq.~\ref{eq:ocp} are diagonal matrices which map the bounded outputs and inputs. The control outputs are limited to match the hardware used ($u_\text{min,max}$) while constraints imposed on the outputs ($y_\text{min,max}$) are used to guarantee safe engine operation as summarized in Table~\ref{tab:mpc_constraints_exp}. In this work the constraints are not overly restrictive, however, in future testing, these limits could be modified to meet specific legislation or design constraints. 

\begin{table}[h!]
	\centering
	\caption{NMPC Constraint Values}\label{tab:mpc_constraints_exp}
	\begin{tabular}{ccc}
		\hline
		Lower bound & Variable & Upper bound \\\hline
		$1\ \mathrm{bar}$ & $\yimep$ & $6\ \mathrm{bar}$ \\
		$0\ \mathrm{CAD aTDC}$ & $\yca$ & $17\ \mathrm{CAD aTDC}$ \\		
		$0\ \mathrm{ppm}$ & $\ynox$ & $500\ \mathrm{ppm}$ \\
		$0\ \mathrm{bar/CAD}$ & $\ymprr$ & $15\ \mathrm{bar/CAD}$ \\
		$0\ \mathrm{ms}$ & $\ufuel$ & $1.50\ \mathrm{ms}$ \\
		$0\ \mathrm{ms}$ & $\uwater$ & $1.00\ \mathrm{ms}$ \\
		$150\ \mathrm{CAD}$ & $\unvo$ & $360\ \mathrm{CAD}$ \\
		\hline
	\end{tabular}
\end{table}

\section{Experimental deployment of LSTM-NMPC}
\begin{figure}[thpb]
	\centering
	\includegraphics[width=0.49\textwidth]
    {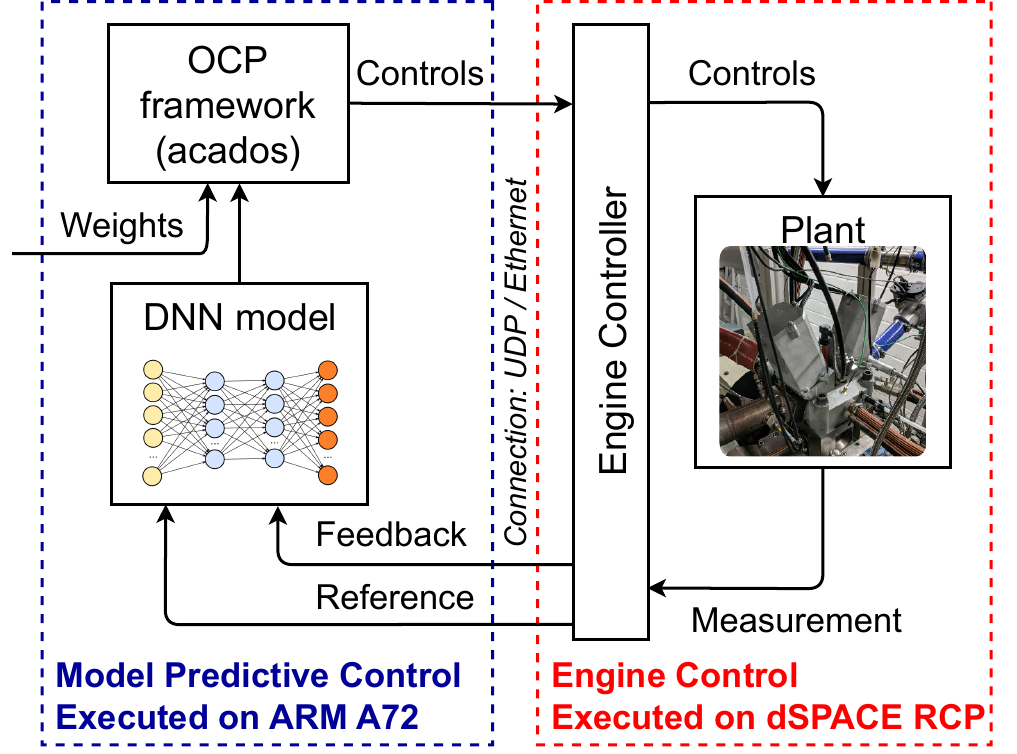}
	\caption{Block diagram of the split controller structure running on a Raspberry Pi 400 and Rapid Control Prototyping (RCP) unit dSPACE MABX II}
	\label{fig:rpimabxsplit}
\end{figure}

One of the main challenges of deploying an NMPC for engine control is the limited calculation time available. For this work, the engine is operated at $1500$~rpm resulting in one complete engine cycle taking $80$~ms. The combustion metrics, calculated on the FPGA, are completed at 60~crank angle degrees~(CAD) after top dead center (aTDC)~\cite{Gordon2018} and the valve timing setpoint must be known by 260~CAD, so only 200~CAD or 22~ms are available for the NMPC calculation. To meet the real-time requirements, the computationally efficient open source package \texttt{acados} is used~\cite{Verschueren2021}. When compared to previous LSTM-based NMPC integration~\cite{GordonEnergies2022}, this work offloads the NMPC calculation to an ARM Cortex~A72 processor located in a Raspberry Pi 400 (RPI) overclocked to 2.2 Ghz clock frequency. The RPI communicates with the main engine controller running on the dSPACE MABX~II over UDP as shown in Figure~\ref{fig:rpimabxsplit}. 

\begin{figure*}[bthp]
	\centering
	\setlength{\figH}{12cm}
	\setlength{\figW}{15cm}
	\input{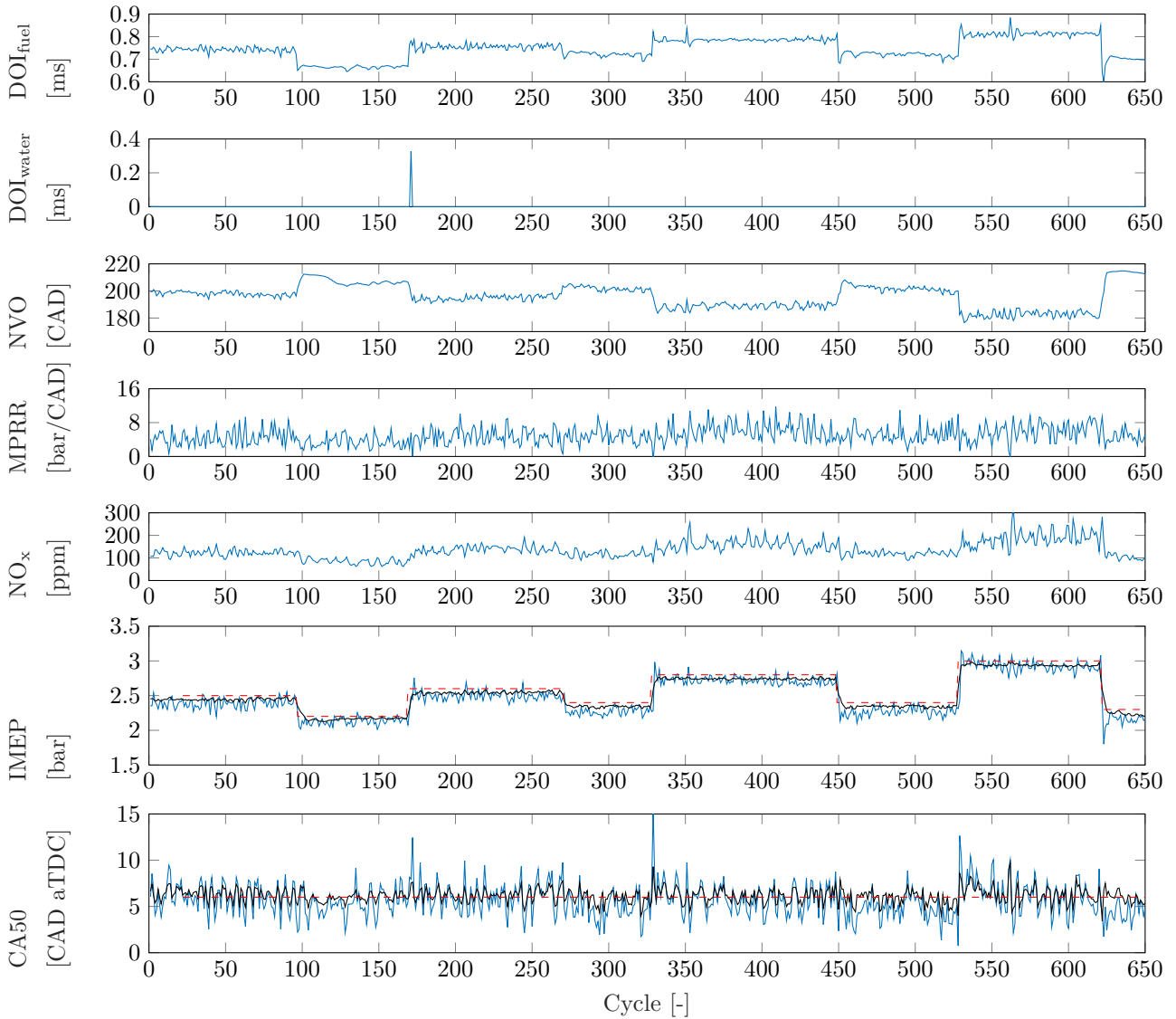}
	\caption{Experimental cycle-to-cycle NMPC implementation on ARM Cortex~A72: Showing IMEP reference tracking performance. Experimental data shown in blue, DNN model data in black and reference shown in dashed red.   }
	\label{fig:RPI_imep_tracking}
\end{figure*}
The NMPC calculation on the RPI is standalone from the main engine controller (dSpace MABX II) other than receiving the current measured states (MPRR, IMEP, CA50 and \nox~emissions) and reference from the MABX as shown in figure~\ref{fig:RPI_imep_tracking}. Both the optimizer and LSTM model are calculated on the RPI where the calculated actuations for the next cycle are sent to the MABX~II. One significant benefit of this formulation is that the NMPC model can be updated or replaced without rebuilding the main engine control software, thus significantly reducing the development time of the controller. Additionally, this modular controller design allows for the NMPC to be executed on any external processor that can interface with the MABX~II over the user datagram protocol (UDP).

When compared to other MPC implementations, in simulation \texttt{acados} outperforms both MATLAB's MPC toolbox using \texttt{fmincon} as well as the \texttt{FORCES PRO} backends~\cite{Norouzi2022}. This difference in solver performance is attributed to the higher dimension of the state than the control input vector in addition to the short prediction horizon of three cycles needed to model HCCI ICE dynamics. This allows the OCP solver to take full advantage of the condensation benefits~\cite{Diehl2009, Frison2016}.

As no discretization is required for the model used, the plant model can be directly implemented using the discrete dynamics interface of \texttt{acados}. The Gauss-Newton approximation is used for the computation of the Hessian in the underlying Sequential Quadratic Programming (SQP) algorithm. The Optimal Control Problem in Eq.~\ref{eq:ocp} leads to a band diagonal structure in the matrices of the Quadratic Problems (QPs) which are solved using the Interior Point (IP) based QP solver \texttt{hpipm}~\cite{frison2020hpipm}, that is provided by the \texttt{acados} package~\cite{Verschueren2021}.

Using a model-in-the-loop (MiL) simulation run on the targeted hardware; the weights, number of SQP iterations and prediction horizon are defined for the NMPC. These MiL simulations showed that three SQP iterations can be completed in the available calculation window. Additionally, as previous work has shown that the cycle-to-cycle dependency of HCCI lasts approximately two cycles so a prediction horizon of three cycles is sufficient~\cite{WoutersJSAE2019}. 

For experimental implementation, a reference load profile is provided to the controller using a target IMEP. Additionally, a reference CA50 is provided to keep the combustion phasing at an efficient operation point of 6~CAD~aTDC. Using the constraints specified in Table~\ref{tab:mpc_constraints_exp} the NMPC is implemented and the experimental LSTM-NMPC performance can be seen in figure~\ref{fig:RPI_imep_tracking}. As the NMPC allows for multi-objective tracking both the IMEP and CA50 tracking performance will be evaluated. A simulated load profile of varying IMEP steps is provided resulting in a tracking performance of 0.133~bar RSME while being able to keep the CA50 at the setpoint of 6~CAD~aTDC with an RSME of 1.83~CAD. As expected, the duration of fuel injection corresponds with the requested IMEP. This trend is also seen in the NVO duration where with increased fuelling a reduced NVO is requested to allow more air into the cylinder.

The constraints applied are obeyed for all but one of the cycles (cycle 552) where the 300~ppm constraint on \nox emissions is slightly exceeded to 305~ppm. This is likely due to plant-model mismatch, resulting from the model predicting a slightly lower \nox output than experimental data. With reference to Figure~\ref{fig:RPI_imep_tracking}, the model prediction for both CA50 and IMEP parameters appear damped compared to the experimental values. However, for both engine performance metrics, the model is able to capture the trends quite accurately with the DNN model predicting IMEP with an RSME of 0.09~bar and for CA50 the RSME is 1.27~CAD. 

The experimental testing of the \texttt{acados} NMPC executed on the ARM Cortex~A72 external processor resulted in an average execution time of 1.18~ms with all calculations taking below 1.4~ms for each of the 650 cycles tested. Even with the 1~ms UDP communication time, the NMPC is below the 22~ms available and shows that if needed the model complexity could be increased with the computational power available. 

\begin{figure}[thpb]
	\centering
	\setlength{\figH}{4cm}
	\setlength{\figW}{7.5cm}
	\input{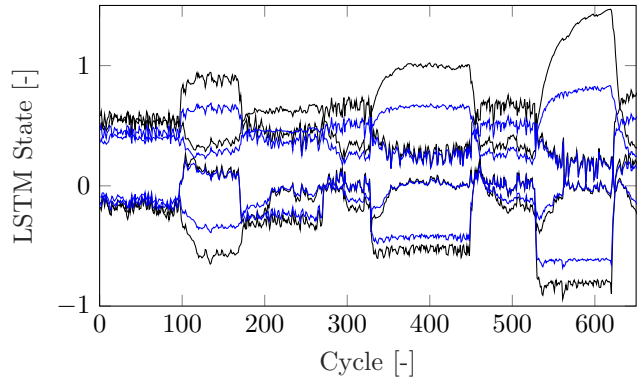}
	\caption{LSTM cell and hidden states during experimental implementation. Cell states in black and hidden states in blue.}
	\label{fig:RPI_lstm_states}
\end{figure}

The LSTM model used in the NMPC implementation has the ability for the cell and hidden states to vary depending on the current engine output. This allows the LSTM model to adapt as the engine changes. The change in the LSTM states can be seen in Figure~\ref{fig:RPI_lstm_states}. These are not physical but rather internal states resulting from the structure of the DNN model used.

When compared to the time and resource-intensive process traditionally used for developing a look-up table-based control strategy the LSTM-based NMPC provides a strategy to allow for rapid controller development. For the model used in this work, the experimental data collected (65,000 engine cycles) took 1.5 hours of testbench time to collect. Often the most significant drawback of blackbox-based models is the required training time for the model itself, however, using modern computing hardware (Intel I7-12700K based PC with a NVIDIA RTX 3090Ti) the model training took just under 3 hours. Taking advantage of the flexibility of \texttt{acados} allows for efficient NMPC design and integration to the external processor. This control design process is significantly quicker than traditional engine control development which can save time and reduce calibration effort and costs while providing improved controller performance by allowing for the integration of an optimal control strategy on experimental hardware.

\section{Conclusions}

Overall, the development and experimental implementation of an LSTM-based NMPC executed on an external processor using the \texttt{acados} framework was shown. To train the DNN process model, which has 2260 learnable parameters, a data set of 65,000 consecutive cycles was used. The DNN network consists of an LSTM layer, used to capture long-term dependencies and cyclic coupling, surrounded by fully connected layers providing a computationally efficient model of the HCCI combustion outputs (IMEP, MPRR, CA50 and NOx). This model resulted in an error below 5\% for all outputs on validation data.

Using this DNN model of the HCCI process, the~open-source software \texttt{acados} provided the required embedded programming for real-time implementation of the LSTM-NMPC. Experimental testing showed HCCI cycle-to-cycle combustion control provided good load and combustion phasing tracking while simultaneously obeying process constraints. With an average turnaround time of 1.18~milliseconds, this work showed the potential of the ML-based NMPC on a real-time system where only 22~ms was available for NMPC calculation and actuation. The advantage of the proposed LSTM-NMPC implementation allows for a shorter controller development time and efficient MPC integration. 

The proof of concept shown in this work can then be further improved by tightening process constraints to match emission legislation and to improve the longevity of the engine. Additionally, the developed toolchain for ML-based NMPC could be applied to other nonlinear constrained systems including ICEs utilizing alternative fuels or hydrogen fuel cells. 

\section*{Acknowledgements}
The research was performed under the Natural Sciences Research Council of Canada Grant 2022-03411 and as part of the Research Unit (Forschungsgruppe) FOR 2401 “Optimization based Multiscale Control for Low Temperature Combustion Engines” which is funded by the German Research Association (Deutsche Forschungsgemeinschaft, DFG).

\footnotesize
\printbibliography

\end{document}